\documentclass[twocolumn]{aastex62}
\usepackage{amsmath}
\usepackage{boxhandler}
\usepackage{threeparttablex}

\graphicspath{{./}{figures/}}

\usepackage{amssymb}
\usepackage{epstopdf, graphicx}

\usepackage{times}

 \newcommand{\gt}{\ensuremath{>}}

\renewcommand{\thefigure}{\arabic{figure}}
\shorttitle{Lensed Quasar at z=6.51}
\shortauthors{Fan et al.}

\begin{document}

\title{The Discovery of a Gravitationally Lensed Quasar at $z=6.51$}

\correspondingauthor{Xiaohui Fan}
\email{fan@as.arizona.edu}

\author[0000-0003-3310-0131]{Xiaohui Fan}
\affiliation{Steward Observatory, University of Arizona, 933 N. Cherry Ave., Tucson, AZ 85721 }

\author[0000-0002-7633-431X]{Feige Wang}
\affiliation{Department of Physics, Broida Hall, University of California, Santa Barbara, CA 93106}

\author[0000-0001-5287-4242]{Jinyi Yang}
\affiliation{Steward Observatory, University of Arizona, 933 N. Cherry Ave., Tucson, AZ 85721 }

\author[0000-0001-6812-2467]{Charles R. Keeton}
\affiliation{Department of Physics and Astronomy, Rutgers University,  Piscataway, NJ 08854}

\author[0000-0002-5367-8021]{Minghao Yue}
\affiliation{Steward Observatory, University of Arizona, 933 N. Cherry Ave., Tucson, AZ 85721 }

\author[0000-0001-6047-8469]{Ann Zabludoff}
\affiliation{Steward Observatory, University of Arizona, 933 N. Cherry Ave., Tucson, AZ 85721 }

\author[0000-0002-1620-0897]{Fuyan Bian}
\affiliation{European Southern Observatory, Vitacura, Santiago 19, Chile}

\author[0000-0002-4236-2339]{Marco Bonaglia}
\affiliation{Osservatorio Astrofisico di Arcetri, Largo Enrico Fermi 5, Florence, Italy}

\author{Iskren Y. Georgiev}
\affiliation{Max Planck Institut f\"ur Astronomie, K\"onigstuhl 17, D-69117, Heidelberg, Germany}

\author[0000-0002-7054-4332]{Joseph F. Hennawi}
\affiliation{Department of Physics, Broida Hall, University of California, Santa Barbara, CA 93106}

\author[0000-0001-6239-3821]{Jiangtao Li}
\affiliation{Department of Astronomy, University of Michigan, 1085 S. University, Ann Arbor, MI 48109}

\author[0000-0002-3461-5228]{Ian D. McGreer}
\affiliation{Steward Observatory, University of Arizona, 933 N. Cherry Ave., Tucson, AZ 85721 }\

\author[0000-0003-3997-5705]{Rohan Naidu}
\affiliation{Center for Astrophysics, 60 Garden Street, Cambridge, MA 02138}

\author[0000-0001-9879-7780]{Fabio Pacucci}
\affiliation{Department of Physics, Yale University, New Haven, CT 06511}

\author{Sebastian Rabien}
\affiliation{Max-Planck-Institut  f\"ur  extraterrestrische Physik, Giessenbachstrasse, 85748 Garching, Germany}

\author{David Thompson}
\affiliation{Large Binocular Telescope Observatory,933 North Cherry Avenue, Tucson, AZ 85721}

\author[0000-0001-9024-8322]{Bram Venemans}
\affiliation{Max Planck Institut f\"ur Astronomie, K\"onigstuhl 17, D-69117, Heidelberg, Germany}

\author[0000-0003-4793-7880]{Fabian Walter}
\affiliation{Max Planck Institut f\"ur Astronomie, K\"onigstuhl 17, D-69117, Heidelberg, Germany}

\author{Ran Wang}
\affiliation{Kavli Institute for Astronomy and Astrophysics, Peking University, Beijing 100871, China}
\affiliation{Department of Astronomy, School of Physics, Peking University, Beijing 100871, China}

\author[0000-0002-7350-6913]{Xue-Bing Wu}
\affiliation{Kavli Institute for Astronomy and Astrophysics, Peking University, Beijing 100871, China}
\affiliation{Department of Astronomy, School of Physics, Peking University, Beijing 100871, China}

\begin{abstract}
Strong gravitational lensing provides a powerful probe of the physical properties of quasars and their host galaxies.
A high fraction of the most luminous high-redshift quasars was predicted to be lensed due to magnification bias.
However, no multiple imaged quasar was found at $z>5$ in previous surveys. 
We report the discovery of J043947.08+163415.7, a strongly lensed quasar at $z=6.51$, the first such object detected at the epoch of reionization, and the brightest quasar yet known at $z>5$. 
High-resolution {\em HST} imaging reveals a multiple imaged system with a maximum image separation $\theta \sim 0.2\arcsec$, best explained by a model of three quasar images lensed by a low luminosity galaxy at $z\sim 0.7$, with a magnification factor of $\sim 50$. 
The existence of this source suggests that a significant population of strongly lensed, high redshift quasars could have been missed by previous surveys, 
as standard color selection techniques would fail when the quasar color is contaminated by the lensing galaxy.
\end{abstract}


\section{Introduction} \label{sec:intro}

Luminous quasars at $z >$ 6 allow detailed studies of the evolution of supermassive black holes (SMBHs) and the intergalactic medium (IGM) at early cosmic times. 
To date, $\sim$ 150 quasars have been discovered at $z >$ 6, with the highest redshift at $z =7.54$  \citep{Banados18}. Detections of such objects indicate the existence of billion solar mass ($M_{\odot}$) SMBHs merely a few hundred million years after the first star formation in the Universe and provide the most stringent constraints on the theory of early SMBH formation \citep{Volonteri12}.

Much of our understanding of the nature of high-redshift quasars assumes that their measured luminosities are intrinsic to the quasars themselves. However gravitational lensing can substantially brighten quasar images. This effect is particularly important in flux-limited surveys, which are sensitive to the brightest sources; the resulting magnification bias \citep{Turner80} could cause a significant overestimation of the SMBH masses powering these objects. A large lensing fraction among the highest redshift luminous quasars has long been predicted \citep{Wyithe02,Comerford02} and was suggested as a solution to the difficulty in forming billion $M_{\odot}$ SMBHs in the early universe. However,
the two highest redshift known lensed quasars are at $z \sim 4.8$ \citep{McGreer10,More2016}, discovered in the Sloan Digital Sky Survey (SDSS); 
no multiple imaged systems were discovered at $0.1\arcsec$ resolution among the more than 200 quasars at $z=4 - 6.4$ observed in two {\em HST} 
programs \citep{Richards06,McGreer14}.
The lack of the high-redshift lensed quasars has been a long-standing puzzle.  %
The solution could be either 
a reduced magnification bias due to a flat quasar luminosity function \citep{Wyithe04} or 
a strong selection effect against lensed objects arising from
the morphology or color criteria  used in quasar surveys \citep{Wyithe02b}.

 \begin{figure*}[h!]
\centering
\includegraphics[width=\linewidth,trim=0cm .7cm 0cm .7cm]{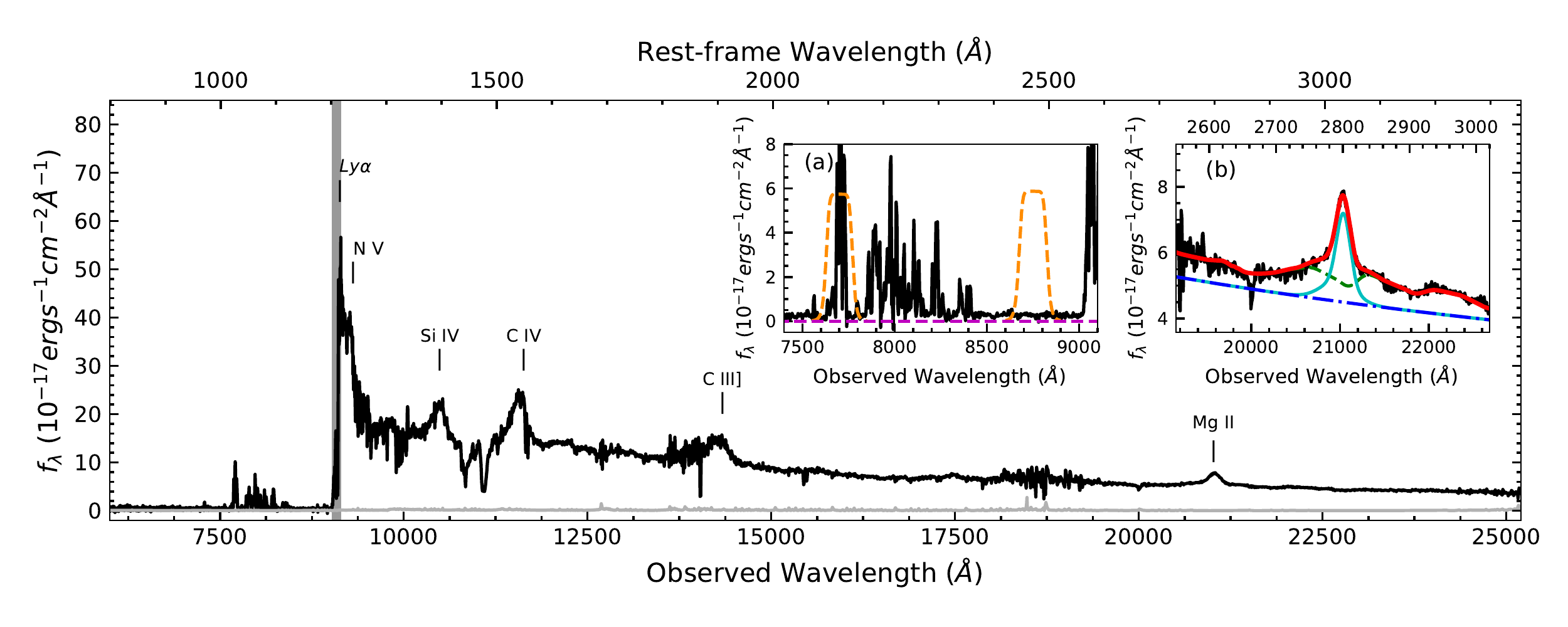}
\caption{\small
\textbf{
Combined optical and near-infrared spectrum of the lensed quasar J0439+1634 at {\em z} = 6.51.} 
The optical portion of the spectrum is from the Binospec instrument on the 6.5m MMT telescope and the LRIS  instrument on the 10m Keck-I telescope. The near-infrared portion of the spectrum is from the GNIRS instrument on the 8.2m Gemini North Telescope. The proximity zone around the quasar is denoted by the grey shaded area blueward of Ly$\alpha$; its size ($R_{p} = 3.61 \pm 0.15 \rm  Mpc$) is $\gt 2\times$ smaller than for other luminous quasars at $z\sim 6.5$ \citep{Wu15,Mazzu17}, suggesting that the intrinsic ionizing flux is much lower. 
{\bf Insert (a)} shows the spectrum in the Ly$\alpha$ forest region. A faint continuum is clearly detected in the darkest region of the quasar Gunn-Peterson trough, suggesting the presence of a foreground galaxy. Orange dashed lines are the traces of the {\em HST}/ACS ramp filters used to image the lensing galaxy and quasar (images shown in Figure 3).
{\bf Insert (b)} shows the Mg II region of the quasar spectrum. The red line is the best fit spectrum when including a power-law continuum, Balmer continuum, and Mg II+Fe II emission. The best-fit redshift based on Mg II is $6.511 \pm 0.003$. The best-fit FWHM of the MgII line is 2924$\pm$188 km s$^{-1}$, yielding a SMBH mass \citep{McLure04} of (4.93 $\pm 0.56) \times 10^9 M_{\odot}$ before correction for lensing magnification.
}
\label{fig:spectrum}
\end{figure*}

\begin{figure*}[h!]
\vspace{-2cm}
\centering
\includegraphics[width=0.7\linewidth,trim=0cm 2.8cm 0cm 3.5cm]{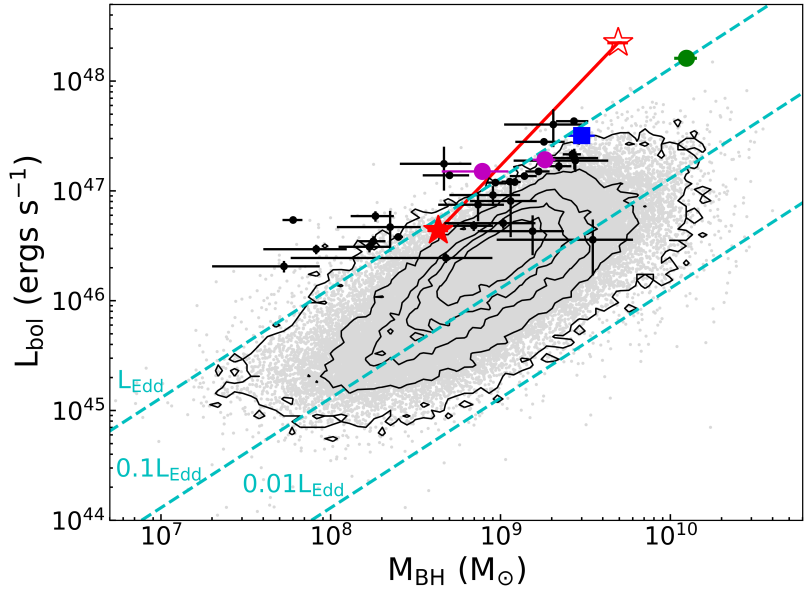}
\vspace{-2cm}
\caption{\small
\textbf{ 
Distribution of quasar bolometric luminosities and SMBH masses estimated from Mg II emission}.
The open red star represents J0439+1634 without lensing correction; the filled red star represents the same object after applying a lensing magnification correction factor of 51$\times$ (from the fiducial lensing model in Table \ref{table:lensmodel}). 
The green circle represents SDSS J0100+2922 at $z=6.30$ \citep{Wu15}, the blue square SDSS J1148+5251 at $z=6.42$ \citep{Fan03}, and the magenta circles ULAS J1120+0641 at $z=7.09$ \citep{Mortlock11} and ULAS J1342+0928 at $z=7.54$ \citep{Banados18}.
Black dots denote other $z\gtrsim6$ quasars \citep{Wu15,Mazzu17}. The black contours and grey dots show SDSS low redshift quasars \citep{shen11} (with broad absorption line quasars excluded). The error bars represent the 1$\sigma$ measurement errors.  For comparison, the dashed lines illustrate fractions of the Eddington luminosity. 
}
\label{fig:bhmass}
\end{figure*}

In our wide-area survey of luminous $z\sim 7$ quasars \citep{Wang17}, we discovered an ultraluminous quasar UHS J043947.08+163415.7 (hereafter J0439+1634) at $z=6.51$.
Subsequent {\em Hubble Space Telescope} ({\em HST}) imaging shows that it is a multiple imaged gravitationally lensed quasar, the most distant strongly lensed quasar yet known.
We present the initial discovery and followup imaging observations that confirm its lensing nature in \S2. 
In \S3, we present the lensing model in detail.  In \S4, 
we discuss the possibility of a large number of high-redshift lensed quasars missed in previous surveys due to bias in color selection. We use a $\Lambda$CDM cosmology with $\Omega_\text{M}=0.3$, 
$\Omega_\Lambda=0.7$ and $H_0=70$ \text{km s}$^{-1}$.

\section{J0439+1634: A Lensed Quasar at z=6.51} \label{sec:obs}

\subsection{Photometric selection and initial spectroscopy}
 J0439+1634 was selected by combining photometric data from the UKIRT Hemisphere Survey (UHS; \cite{UHS}) in the near-infrared J band, the Pan-STARRS1 survey (PS-1; \cite{PS1}) at optical wavelengths, and the {\em Wide-field Infrared Survey Explorer} ({\em WISE}; \cite{wright10}) archive in the mid-infrared. It was chosen as a high-redshift quasar candidate based on it having a $z$-band dropout signature with  $z_{AB}=19.49\pm0.02$, $y_{\rm Vega} =17.63 \pm 0.01$, and a red $z_{AB} - y_{AB} = 1.86 \pm 0.02$, 
along with a blue power-law continuum ($J_{\rm Vega}= 16.52 \pm 0.01$, $y_{\rm AB}-J_{\rm Vega}= 1.11\pm0.02$), and a photometric redshift of $z\sim 6.5$. The object has a weak $i$-band detection in PS1 ($i_{\rm AB} = 21.71 \pm 0.05$), but  
is strongly detected in all bands in the Two Micron All Sky Survey (2MASS; \cite{strutskie06}), at $J_{\rm Vega}= 16.48 \pm 0.12$, 
 $H_{\rm Vega}= 15.96 \pm 0.17$, and  $J_{\rm Vega}= 15.06 \pm 0.13$, respectively, as well in 
all four {\em WISE} bands, with Vega magnitudes of $13.98 \pm 0.03, 13.24 \pm 0.03, 10.28 \pm 0.08$ and $7.17 \pm 0.13$, respectively, from W1 to W4. 

The initial identification spectrum, obtained on 6 February, 2018, with the Binospec optical spectrograph \citep{BINOSPEC} on the 6.5m MMT telescope,
shows a prominent break consistent with a strong Ly$\alpha$ line at $z\sim 6.5$. 
Follow-up optical and near-infrared spectra were acquired with MMT/Binospec, 
the Low Resolution Imaging Spectrograph (LRIS, \citealt{LRIS}) on the 10m Keck-I Telescope,
and the GNIRS instrument \citep{GNIRS} on the 8.2m Gemini-North Telescope.
The combined optical-IR spectrum is shown in Figure 1. Strong MgII emission is detected by GNIRS, yielding a redshift of $z=6.511\pm0.003$. 

J0439+1634 is roughly 40\% brighter than the luminous $z=6.30$ quasar SDSS J0100+2802 \citep{Wu15},
making it the brightest quasar known at $z>5$. It is also the brightest submm quasar at $z>5$; it is detected by the SCUBA-2 instrument \citep{SCUBA} on the James Clerk Maxwell Telescope (JCMT) with a total flux of 26.2$\pm$1.7 mJy at 850 $\mu$m.
However, its high luminosity is likely not intrinsic, but instead boosted via gravitational lensing. 
The optical spectrum of J0439+1634 shows a faint, continuous trace at $\lambda < 9000$ \AA, visible in the middle of the deepest region of quasar Gunn-Peterson absorption at 8500\AA $<$ 9000\AA ($z_{\rm abs} > 6$). This trace
extends beyond the quasar Lyman Limit at $\lambda < 6840$ \AA, blueward of the IGM transmisssion spikes in the quasar Ly$\beta$ region. No quasar continuum transmission is expected at these wavelengths due to the extremely high IGM optical depth \citep{Fan06},
indicating the presence of a foreground object within the $1\arcsec$ spectroscopic slit.
The lensing hypothesis is further supported by the presence of a very small quasar proximity zone 
(Figure \ref{fig:spectrum})
and an apparent super-Eddington accretion rate based on the Mg\,{\sc ii} measured SMBH mass (Figure \ref{fig:bhmass}), both of which can be explained with a significant lensing magnification.

\subsection{High Resolution Imaging}

\begin{figure*}[h!]
\centering
\includegraphics[width=0.7\linewidth]{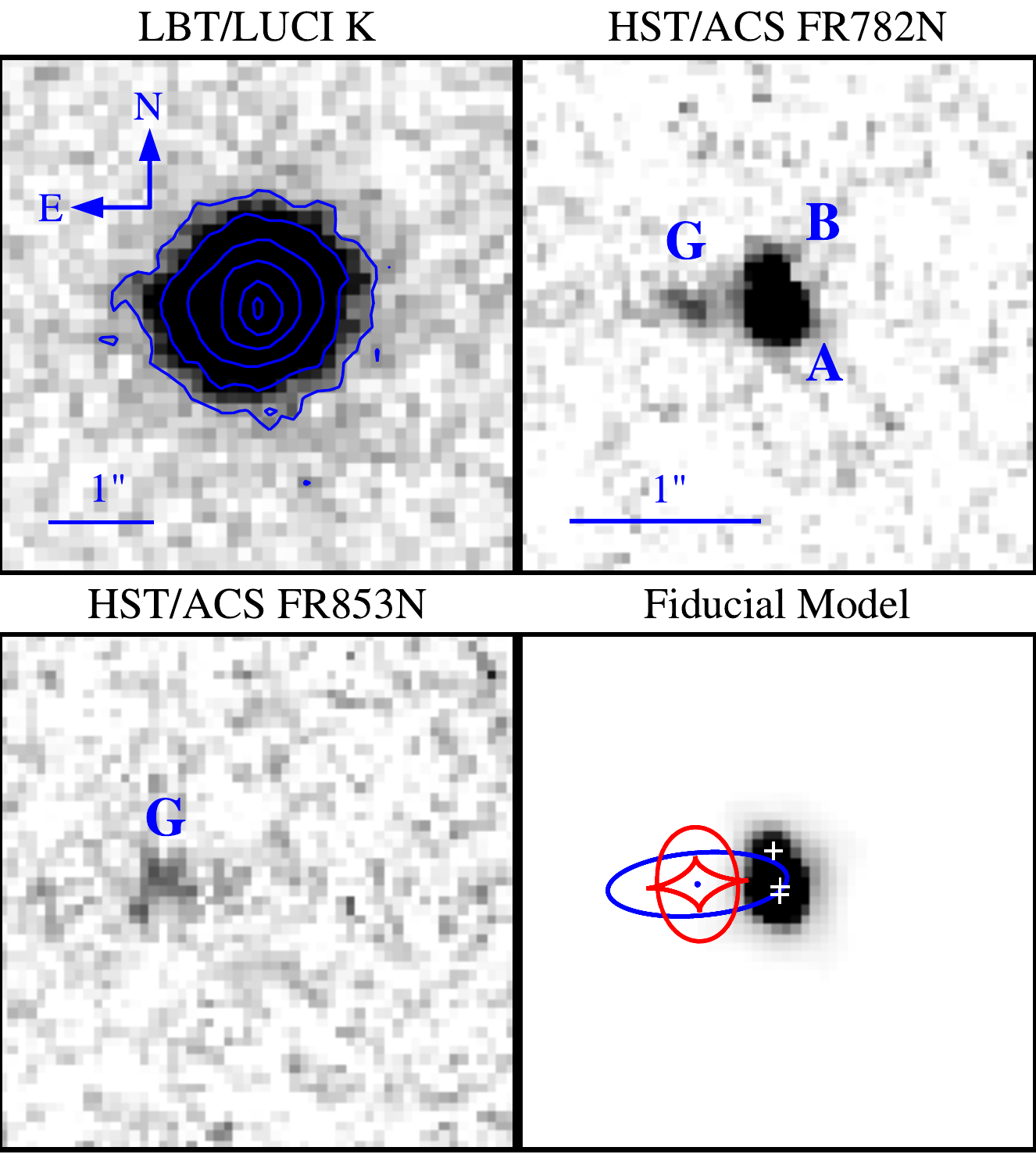}
%
\caption{\small
\textbf{High-resolution images of the strongly lensed quasar J0439+1634 and the best-fit three-image lensing model.}
{\bf A:}
LBT/ARGOS LUCI image in the K-band. With ground layer AO correction, the FWHM of the PSF is $0.24\arcsec$. The quasar image has a FWHM of $0.30''\pm 0.01''$. The contours show the image core elongated in the North-South direction as well as excess light towards the East, consistent with the high resolution {\em HST} imaging.  
{\bf B:} {\em HST}/ACS WFC image with the FR782N ramp filter centered at 7700\AA, covering the quasar Ly$\beta$ emission. This ``galaxy+quasar'' image is resolved into at least two point-like components (A and B) and a faint extended source towards the East (G). 
{\bf C:} {\em HST}/ACS WFC image with the FR853N ramp filter centered at 8750\AA, covering the deepest part of the quasar Gunn-Peterson trough. This ``galaxy only'' image is used to determine the location and shape parameters of the lensing galaxy. 
{\bf D:}  Best-fit three image lensing model to the {\em HST}/ACS FR782N image, using the lens location and shape derived from the FR853N image.
White crosses show the locations of the best-fit quasar images and blue lines show lensing critical curves of the fiducial lensing model.
Red lines show the lensing caustics in the source plane.
In this model, the total magnification is $51.3\pm1.4$ and the Einstein radius is $0.17\arcsec$, which corresponds to a circular velocity of $v_c = 160_{-6}^{+8}$ km s$^{-1}$ assuming a lens redshift $z = 0.67^{+0.19}_{-0.15}$.
}
\label{fig:image+lensmodel}
\end{figure*}

J0439+1634 appears as an unresolved point source on 
 archival PS1 and UHS images (seeing of $\sim 1.5''$) and on deeper near-infrared images taken with the Fourstar instrument \citep{FOURSTAR} on the 6.5m Magellan-1 Telescope (seeing $\sim 0.8''$).
To test the lensing hypothesis,  we obtained a high resolution K-band image using the Advanced Rayleigh guided Ground layer adaptive Optics System (ARGOS; \citealt{ARGOS})
on the $2\times 8.4$m Large Binocular Telescope, with a ground-layer AO corrected FWHM of $0.24\arcsec$. 
This image (Figure \ref{fig:image+lensmodel}A), taken with the LUCI \citep{LUCI} instrument, 
marginally resolves J0439+1634 beyond the PSF (FWHM = $0.30''\pm 0.01''$).

Even more revealing are the high resolution observations of J0439+1634 with 
the Advanced Camera for Surveys (ACS)
on the {\em HST}, taken on 3 April, 2018, using two intermediate band ($\Delta \lambda \sim 200$\AA) ramp filters (Figure \ref{fig:spectrum}).
The FR782N observation is centered at 7700\AA, fully covers the quasar Ly$\beta$ emission, and is the shortest wavelength at which quasar emission is still detectable, thus providing the highest possible spatial resolution of $0.075\arcsec$. 
The FR853N observation is centered at 8750\AA, within the Gunn-Peterson trough, and images only the foreground galaxy.
The ``galaxy+quasar'' FR782N image (Figure \ref{fig:image+lensmodel}B) clearly resolves the system into multiple components:  there are at least two point sources separated by $0.2\arcsec$ and a faint, extended source $\sim 0.5\arcsec$ to the east, which we interpret as the lensing galaxy. 
The ``galaxy-only'' FR853N image (Figure \ref{fig:image+lensmodel}C) shows only the lensing galaxy, best fit with an exponential profile, an ellipticity of $\sim 0.65$, and an effective radius of $\sim 0.4\arcsec$. 

\subsection{Properties of the lensing galaxy}
We use the best-fit galaxy position and shape parameters from the FR853N image to derive the lensing galaxy flux in the two {\em HST} bands and LBT K-band:
$AB_{7700 \AA} = 22.40 \pm 0.05$, $AB_{8750 \AA} = 22.07 \pm 0.07$ and $K_{\rm Vega} = 18.86 \pm 0.19$.
The non-detection in the blue channel of the Keck/LRIS spectrum yields an upper limit of $g_{\rm AB} >24$ for the galaxy. 
We estimate the synthetic PS-1 $g, r, i$ band magnitudes of the lensing galaxy using the spectrum of J0439+1634 (Figure \ref{fig:spectrum}), which shows the trace of the lensing galaxy spectrum in the quasar Gunn-Peterson trough. 
We scale the spectrum by matching it to the {\em HST}/FR853N band magnitude, which does not include quasar flux.
We choose a wavelength range free of quasar flux, between 8600 and 8900 \AA\ in the Gunn-Peterson trough, and blueward of the Lyman limit ($<$ 6840 \AA), to calculate the magnitudes. 
For the spectrum between 6840 \AA\ and 8600 \AA, we interpolate the continuum by fitting the blue- and red-side spectrum with a spline function.
The synthetic $g$, $r$ and $i$ band AB magnitudes are estimated to be $25.00 \pm 0.90$, $23.29 \pm 0.29$, $22.47 \pm 0.11$, respectively. 

Based on these photometric data and after applying the Galactic extinction correction \citep{Cardelli1989}, we estimate the photometric redshift using the EAZY \citep{Brammer2008} code. The peak value of the p(z) probability distribution is z\_peak = 0.67, and the 1 $\sigma$ confidence interval from the probability distribution is $0.52 \le z \le 0.86$. 
With the $Le Phare$ code \citep{Arnouts2002,Ilbert2006} and a set of 12 template galaxies using \cite{Bruzual2003} models, we find a best-fit stellar mass of 10$^{9.8}$ $M_{\odot}$.
Deeper photometry is needed to further improve the photometric redshift and stellar mass determinations.

\begin{figure*}[ht!]
\centering
\includegraphics[width=0.85\linewidth]{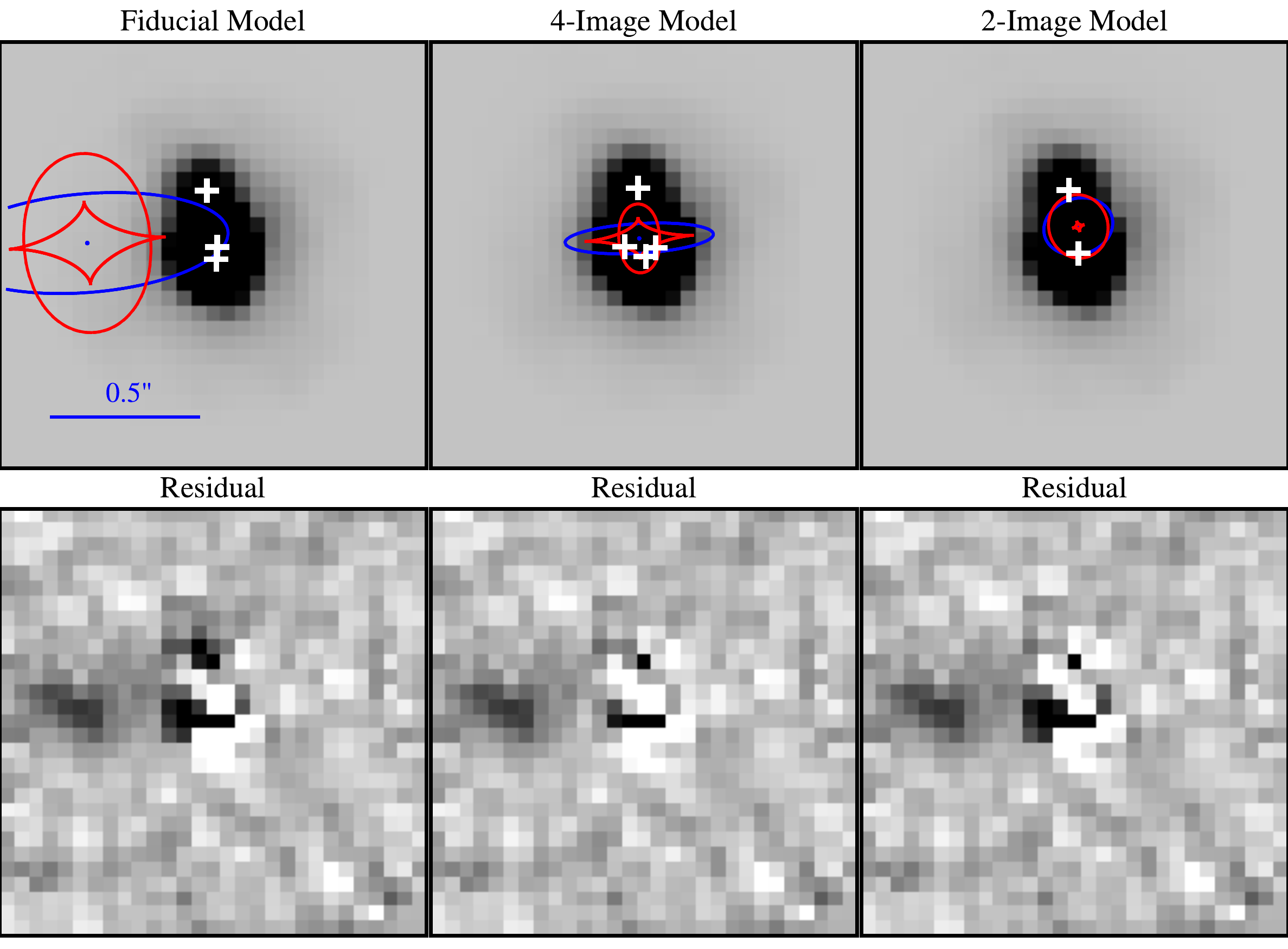}
\caption{\small 
\textbf{Fiducial and Alternative Lens Models: Fits (top row) and residuals (bottom row) of the {\em HST}/ACS FR782N image.}
As in Figure \ref{fig:image+lensmodel}, white crosses are locations of quasar images, and blue and red lines represent the lensing critical curves and caustics, respectively.
}
\label{fig:alternatemodel}
\end{figure*}

\section{Lensing Model}

A purely photometric fit of the {\em HST}/ACS FR782N data using only two quasar images has a significant residual, suggesting a more complex lensing configuration. 
We fit a singular isothermal ellipsoid lensing model, fixing the lens position and ellipticity ($e = 0.65$) to match the observed galaxy in the FR853N image, while varying the lens mass and position angle along with the source position to reproduce the observed configuration \citep{Keeton2001}.
We vary the Einstein ring radius and position angle of the galaxy along with the position of the source.
For each set of parameters, we solve the lens equation to predict the positions of the images, place copies of the {\em HST} PSF at those positions, and compare with the FR782N image to compute a $\chi^2$ goodness of fit.
We then use Markov Chain Monte Carlo methods to sample the parameter space.
The resulting model is depicted in Figure \ref{fig:image+lensmodel} and the parameters are summarized in Table \ref{table:lensmodel}.
To interpret the Einstein radius, we assume the galaxy is a thin rotating disk such that the projected ellipticity reflects the inclination, and we compute the corresponding circular velocity \citep{Keeton1998}.
A three-image model is preferred (Figure \ref{fig:image+lensmodel}D), with a best-fit Einstein radius of $\theta_E = 0.17\arcsec \pm 0.01\arcsec$, corresponding to a circular velocity of $v_c = 160_{-6}^{+8}\ \rm km s^{-1}$ and a high total magnification of $51.3 \pm 1.4$. 
In this model, the separation of the two brighter images is only $0.04\arcsec$, unresolved even by {\em HST}.

We estimate the {\em observed} optical luminosity
at rest-frame 3000\AA\ to be $(4.35 \pm 0.09) \times 10^{47}~{\rm erg~s}^{-1}$ by fitting the calibrated spectrum.
Applying an empirical factor \citep{shen11} to convert the luminosity at 3000\AA\ to the bolometric luminosity gives $L_{\rm bol}=  2.24 \times 10^{48} {\rm ergs~s}^{-1} =  5.85 \times 10^{14} L_{\odot}$.
After correction for magnification factor of 51.3, the bolometric luminosity of J0439+1634 is reduced to  $1.14 \times 10^{13} L_{\odot}$, and  the SMBH mass to $\rm 4.29 \pm 0.60 \times 10^{8} M_{\odot}$. This corresponds to an Eddington ratio of $0.83\pm 0.12$.

However, this model seems to underpredict the flux of the faintest quasar image.
It is not clear whether the discrepancy is due to limitations in the current data (e.g., in the {\em HST} PSF model) or to fundamental problems with this class of lens models.
As an alternative, we consider the possibility that the lens galaxy could actually lie between the quasar images and be blended with them.
In this scenario, the galaxy light detected in the {\em HST} image could be offset from the mass centroid, due perhaps to strong dust obscuration.
For example, if the lensing galaxy is seen mostly edge-on, then we might have detected only the part of the galaxy with the highest surface brightness along the disk.
The smallest residuals are obtained for a highly inclined galaxy with projected ellipticity $e=0.8$, which produces four images and a total magnification of $10.4 \pm 0.2$ (see Figure \ref{fig:alternatemodel} and Table \ref{table:lensmodel}).
The implied circular velocity $v_c = 88_{-3}^{+4}$ km s$^{-1}$ is quite low, comparable to that of the Large Magellanic Cloud.
The orientation is consistent with the hypothesis that the observed galaxy light is from part of the disk.
It also possible that the nearby galaxy is not related to the lensing.
In this case, the true lens galaxy is too faint for detection here, 
could lie between the quasar images, and be relatively round.
We therefore test a third model with ellipticity $e=0.2$, which produces just two images that have a total magnification of ${23.1}_{-0.8}^{+1.4}$.
This model has a modest circular velocity of $v_c = 121_{-4}^{+6}$ km s$^{-1}$.


We consider the fiducial three-image model to be the most likely lensing configuration, because it naturally places the center of the lensing galaxy at the position of the detected galaxy image in the two {\em HST} bands. 
However, further observations are needed to clearly distinguish between the different models.
Images that are deeper than the current {\em HST} observation could fully characterize the lensing galaxy, while observations with higher spatial resolution (possible only with {\em JWST} or ALMA) would reveal whether there are two, three, or four image components.

\section{Discussion}

The probability that a luminous quasars is gravitationally lensed with magnification factor $\mu >2$ at $z \sim 6$ ranges from $\sim 4\%$, if the bright end of the quasar luminosity function is $\Phi(L) \propto L^{-2.8}$ \citep{Jiang16}, to $\sim 20\%$, if the quasar luminosity function is as steep as $\Phi(L) \propto L^{-3.6}$ \citep{Yang16}. 
Yet J0439+1634 is the first strongly lensed quasar discovered at $z>5$ among the several hundred quasars known at this redshift.
A reexamination of the color selection used in previous high-redshift quasar surveys suggests a strong selection bias against lensed quasars. 

Selecting $z\gtrsim 6$ quasars requires either a non-detection \citep{Wang17,Jiang16} or a strong drop in the dropout band below the quasar Lyman break \citep{Mazzu17}.
The presence of a lensing galaxy, however, introduces flux into the dropout bands when the image is not fully resolved. Most lensing galaxies are expected to be massive galaxies at $z\sim 0.5$ - 1.5 and to have detectable $r$- or $i$-band flux in the SDSS or PS-1 survey. For example, among the 62 lensed $z<4$ quasars in the SDSS sample \citep{Inada12} with measurements of the lensing galaxy, the faintest lens has $i_{AB}=21.64$. 
On the other hand, the J0439+1634 lens is among the faintest lensing galaxies known, with $i_{AB} = 22.47$. The faintness of this lens, combined with the high apparent luminosity of the lensed quasar, minimizes the impact of lensing galaxy flux to the overall unresolved quasar+lens color used in candidate selection. 
If the lens were brighter by even 0.5 mag, J0439+1634
would not have been selected as a high-redshift quasar candidate by our color selection criteria \citep{Wang17},
suggesting that previous surveys have potentially missed the majority of lensed quasars at the highest redshifts due to their stringent dropout criteria.
Thus a full modeling of quasar+lens colors and selection procedure modifications are needed to cover the majority of the high-redshift lensed quasar population.

A statistical study of strong lensing properties using the Millennium Simulation \citep{Hilbert} shows that for a source at $z=5.7$, only 5\% of the lensing optical depth is provided by galaxies with a halo mass lower than $7 \times 10^{11} M_{\odot}$, comparable to J0439+1634's lensing galaxy ($v_c = 160$ km s$^{-1}$) in the fiducial three-image lensing model. This implies up to $\sim 20$ lensed high-redshift quasars could have been missed in our survey due to contamination from lensing galaxy light. 
If these lensed quasars do exist, it would significantly impact the measurement of the quasar luminosity function, especially at the brightest end \citep{Wyithe02b}. 
Benefiting  from the boosted flux, an object such as J0439+1634 is a powerful probe of the physical properties
of quasars and their host galaxies as well as serving as an ideal background source for studying high redshift metal absorption lines and early IGM chemical enrichment.


\acknowledgements
We acknowledge the support of the staff at the MMT, Magellan, LBT, Keck and Gemini Telescopes, and thank the Directors of LBTO, Gemini Observatory, JCMT, and STScI for granting us Director Discretionary time for follow up observations of this object. 
 X.F., J.Y., M.Y. and I.D.M. acknowledge support from US NSF Grant AST-1515115, NASA ADAP Grant NNX17AF28G and   HST-GO-13644 grant from the Space Telescope Science Institute. 
 C.K. acknowledges support from US NSF grant AST-1716585.
 A.I.Z. acknowledges support from NSF grant AST-1211874.
 F.P. acknowledges support from the NASA Chandra award No. AR8-19021A and from the Yale Keck program No. Y144.
 B.P.V. and F.W. acknowledge funding through the ERC grants ``Cosmic Dawn'' and ``Cosmic Gas''.
 R.W. and X.-B.W acknowledge support from NSFC grant No. 11533001.
\facilities{UHS, WISE, PS1, MMT, Magellan, LBT, Gemini, Keck, JCMT, HST}



\begin{table}[h!]
\small
\centering
\caption{\textbf{Lens Model Parameters}}

\smallskip
\begin{tabular}{cccc}
\hline
&
Fiducial 3-image model &
Alternate 4-image model &
Alternate 2-image model
\\
\hline
Image 1 &
$(\Delta\mbox{RA},\Delta\mbox{Dec})\equiv(0,0)$  &
$(0,0)$  &
$(0,0)$ 
\\
 & $\mu={5.4}\pm{0.1}$ & $\mu={1.4}$ & $\mu={3.9}_{-0.1}^{+0.3}$ \\ \hline
Image 2 &
$(-0.032, -0.233)$ &
$(-0.027, -0.233)$ &
$(-0.033, -0.215)$ 
\\
 & $\mu={21.8}\pm{0.7}$ & $\mu={5.1}\pm{0.1}$  & $\mu={-19.3}_{-1.2}^{+0.8}$ \\ \hline
Image 3 &
$(-0.035, -0.192)$  &
$(-0.060, -0.203)$ &
--- 
\\
 & $\mu={-24.2}\pm{0.7}$ & $\mu={-2.7}\pm{0.1}$  & \\ \hline
Image 4 &
--- & 
$(0.045, -0.200)$ &
---
\\
 & &  $\mu={-1.2}\pm{0.1}$ & \\ \hline
Source &
$(0.215, 0.076)$  &
$(-0.005, -0.118)$  &
$(-0.025, -0.107)$ 
\\
 &  $\mu_{\rm tot}={51.3}\pm{1.4}$ & $\mu_{\rm tot}={10.4}\pm{0.2}$ &  $\mu_{\rm tot}={23.1}_{-0.8}^{+1.4}$ \\ \hline
Lens &
$(0.438, 0.055)$ &
$(-0.004, -0.171)$ &
$(-0.028, -0.125)$
\\
&
$\theta_E={0.168}\pm{0.001}''$ &
$\theta_E={0.051}\pm{0.001}''$ &
$\theta_E={0.095}\pm{0.001}''$
\\
&
$v_c=160_{-6}^{+8}$ km s$^{-1}$ &
$v_c=88_{-3}^{+4}$ km s$^{-1}$ &
$v_c=121_{-4}^{+6}$ km s$^{-1}$
\\
&
$e=0.65$, $\mbox{PA}={103.1}\pm{0.1}$ &
$e=0.8$, $\mbox{PA}={101.8}\pm{0.6}$ &
$e=0.2$, $\mbox{PA}={112.8}_{-7.5}^{+6.0}$
\\
\hline
\end{tabular}
\label{table:lensmodel}
\end{table}

\end{document}